\documentclass{aa}  
\usepackage{booktabs}
\usepackage{graphicx}

\usepackage{txfonts}
\usepackage{gensymb}
\usepackage{xcolor}
\begin{document}

   \title{Slow solar wind sources. High-resolution observations with a quadrature view}

\titlerunning{Slow solar wind sources. High-resolution observations with a quadrature view}
\authorrunning{Barczynski et al.}

   \author{Krzysztof Barczynski
          \inst{1,2},
          Louise Harra\inst{2,1},
          Conrad Schwanitz\inst{1, 2}, Nils Janitzek\inst{1,2}, 
          David Berghmans \inst{3},
Frédéric Auchère\inst{4},
Regina Aznar Cuadrado\inst{5},
Éric Buchlin\inst{4},
Emil Kraaikamp\inst{3},
David M.\ Long\inst{6,7},
Sudip Mandal\inst{5},
Susanna Parenti\inst{4},
Hardi Peter\inst{5},
Luciano Rodriguez\inst{3},
Udo Sch\"uhle\inst{5},
Phil Smith\inst{6},
Luca Teriaca\inst{5},
Cis Verbeeck\inst{3},
Andrei~N.~Zhukov\inst{3,8},
          }

\institute{ETH-Zurich, H\"onggerberg campus, HIT building, Wolfgang-Pauli-Str. 27, 8093 Z\"urich, Switzerland
\\
              \email{krzysztof.barczynski@pmodwrc.ch}
         \and
      PMOD/WRC, Dorfstrasse 33, CH-7260 Davos Dorf, Switzerland
\and
Solar--Terrestrial Centre of Excellence -- SIDC, Royal Observatory of Belgium, Ringlaan -3- Av. Circulaire, 1180 Brussels, Belgium 
\and
Université Paris-Saclay, CNRS, Institut d’Astrophysique Spatiale, 91405, Orsay, France
\and
Max Planck Institute for Solar System Research, Justus-von-Liebig-Weg 3, 37077 G\"ottingen, Germany
\and
UCL-Mullard Space Science Laboratory, Holmbury St.\ Mary, Dorking, Surrey, RH5 6NT, UK 
\and
Astrophysics Research Centre, School of Mathematics and Physics, Queen’s University Belfast, University Road, Belfast, BT7 1NN, Northern Ireland, UK
\and
Skobeltsyn Institute of Nuclear Physics, Moscow State University, 119992 Moscow, Russia
      }

   \date{Received xxx; accepted xxx}

 
  \abstract
   {The origin of the slow solar wind is still an open issue. One possibility that has been suggested is that upflows at the edge of an active region can contribute to the slow solar wind. }
   {We aim to explain how the plasma upflows are generated,  which mechanisms are responsible for them, and what the upflow region topology looks like.}
   {We investigated an upflow region using imaging data with the unprecedented temporal (3\,s) and spatial (2 pixels = 236\,km) resolution that were obtained on 30 March 2022 with the 174\,\AA~channel of the Extreme-Ultraviolet Imager (EUI)/High Resolution Imager (HRI) on board Solar Orbiter. During this time,  the EUI and Earth-orbiting satellites (Solar Dynamics Observatory, Hinode, and the Interface Region Imaging Spectrograph, IRIS) were located in quadrature ($\sim92\degree$), which provides a stereoscopic view with high resolution. We used the Hinode/EIS (\ion{Fe}{xii}) spectroscopic data to find coronal upflow regions in the active region. The IRIS slit-jaw imager provides a high-resolution view of the transition region and chromosphere. }
   {For the first time, we have data that provide a quadrature view of a coronal upflow region with high spatial resolution. We found extended loops rooted in a coronal upflow region. Plasma upflows at the footpoints of extended loops determined spectroscopically through the Doppler shift are similar to the apparent upward motions seen through imaging in quadrature. The dynamics of small-scale structures in the upflow region can be used to identify two mechanisms of the plasma upflow: Mechanism I is reconnection of the hot coronal loops with open magnetic field lines in the solar corona, and mechanism II is reconnection of the small chromospheric loops with open magnetic field lines in the chromosphere or transition region. We identified the locations in which mechanisms I and II work.}
   {}


   \maketitle
%

\section{Introduction}

The solar wind is a continuous stream of charged particles emitted by the Sun \citep{Parker1958, McComas2003}.
The solar wind is key to understanding the maintenance and dynamics of the heliosphere, in which the planets exist. 
\citet{McComas2003} discussed the three-dimensional solar wind around solar maximum.
The solar wind has two components.
The fast solar wind (faster than 500\,km\,s$^{-1}$ at 1 AU) originates from coronal holes \citep{Temmer2021}.
There are many potential sources of the slow solar wind (slower than 400\,km\,s$^{-1}$ at 1 AU), which are debated. The question of the sources of the slow solar wind is highly complex. The sources are located around the activity belt of the Sun \citep{Abbo2016}.

Doppler velocity maps of active regions show that
the upflows at the border of active regions are observed at coronal temperatures \citep{Sakao2007, Doschek2007, Doschek2008, DelZanna2008, Hara2008, Harra2008}.
The upflow regions are suggested as potential sources of the slow solar wind \citep{Kojima1999, Sakao2007, Harra2008}, and they form  during the early phase of the active region formation and exist throughout the active region lifetime \citep{Brooks2021}. They are always observed when an active region is observed spectroscopically in the corona. 

\citet{Barczynski2021} suggested that at least three parallel mechanisms create the upflow region: (I)  open magnetic field lines can reconnect with the closed hot loops in the lower corona or upper transition region, (II)  open magnetic field lines can reconnect with small-scale chromospheric loops, and (III) the chromospheric
plasma can escape to the solar corona as a result of the expansion of the open magnetic field lines from the photosphere to the corona due to the waves in the solar atmosphere. The role of waves in the upflow was discussed by \citet{Tian2011}.
As a result of the magnetic field reconnection (mechanisms I and II), the magnetic energy is released in the solar atmosphere.
This process heats the surrounding plasma and plasma flows. 
The blueshift is observed above the location of reconnection, and redshifts are observed below the location of reconnection in the solar atmosphere.
The reconnection should leave an imprint on the coronal intensity change with time.
The waves in the lower solar atmosphere might also cause the plasma upflow (mechanism III).

Previous studies of upflow regions \citep{Sakao2007, Doschek2007, Doschek2008, DelZanna2008, Hara2008, Harra2008, Boutry2012, Brooks2021, Barczynski2021} used observations obtained
with Earth-orbiting satellites, that is, with data that were obtained\ from a single viewpoint.
Two methods enable a stereoscopic view of the upflow region:
(1) tracking the upflow region for a few days while the upflow rotates with the Sun \citep{Demoulin2013}, and (2) using simultaneous observations from two satellites that are located with some angular separation, such as\ the two Solar Terrestrial Relations Observatory (STEREO) satellites \citep[e.g.][]{Aschwanden2011}.
Both methods allow studying the long-term changes of the upflow region with the characteristic scale of a few days (method 1) and cadence of STEREO observations, which is at least 2.5 min (method 2).

The new generation of satellites provides a unique opportunity of high spatial and high temporal observation of the solar atmosphere.
The Atmospheric Imaging Assembly \citep[AIA;] []{Lemen2012} on board the Solar Dynamics Observatory \citep[SDO;][]{Pesnell2012} observes the solar disk with a high spatial resolution of 2 pixels, equal to 1.2", which at a distance of 1.0 AU translates into 2 pixels, equal to 872\,km.
The AIA standard time cadence is 12\,s.
The combination of the Solar Orbiter mission \citep{Muller2020} launched in 2020 and Earth-orbiting observatories provides an opportunity for quadrature observations of the Sun when Solar Orbiter is in the appropriate parts of its elliptical orbit. 
The Extreme Ultraviolet Imager \citep[EUI;][]{Rochus2020} on board Solar Orbiter observes the Sun with a Full Sun Imager (FSI) and with two High Resolution Imagers (HRI).
In the 174\,\AA~channel, the HRI telescope (known as HRI$_{EUV}$) provides a spatial resolution of 2 pixels, equal to 1", which at a distance of 0.33 AU translates into an unprecedented spatial resolution of 2 pixels, equal to 236.8 km.
The HRI temporal resolution is 3\,s.
For the first time, we show a quadrature observation of a coronal upflow region with high spatial and high temporal resolution data.

We use a combination of Solar Orbiter HRI$_{EUV}$ data with Interface Region Imaging Spectrograph  \citep[IRIS;][]{DePontieu2014} and Hinode EUV Imaging Spectrometer \citep[Hinode/EIS;][]{Culhane2007}  observations (Section~\ref{sect:observation}). This provides a view of the upflow footpoints from the chromosphere to the corona, complementing the imaging data that provide details of the structure emanating from the upflow region. This is hard to see in imaging because the corona is optically thin when the Sun is observed face on. 
Using Doppler imaging, we compare the line-of-sight plasma velocity in the upflow region and the apparent motions of the coronal plasma measured near quadrature over the plane of sky  along the extended loops that are rooted in the upflow region. (Section~\ref{sect:topology}).
In Section~\ref{sect:dynamics} we discuss the relation between the coronal emission and the underlying transition region and chromosphere.
In particular, we focus on the dynamics of the small-scale features observed in the transition region and chromosphere.
In Section~\ref{sect:discussion} we discuss the properties of the coronal upflow region obtained from the quadrature observation and the plasma dynamics in the solar corona and in the underlying part of the solar atmosphere.
We conclude with our results in Section~\ref{sect:conclusion}.

\section{Observations and data processing}\label{sect:observation}

\subsection{EUI observations}
We analysed images of the solar atmosphere obtained with the EUI on board Solar Orbiter.
The EUI consists of three telescopes: the dual-band FSI, which observes the full solar disk at 174\,\AA\ and 304\,\AA, and the two HRIs, which show the solar atmosphere in the Lyman-$\alpha$ line at 1216\AA\ (HRI$_{Ly-\alpha}$) and in the extreme ultraviolet (EUV) at 174\,\AA\ (HRI$_{EUV}$).
We used level-2 data from HRI$_{EUV}$
showing the upper transition region and lower coronal emission (1\,MK), mainly originating from the \ion{Fe}{ix} and \ion{Fe}{x} lines.

We analysed the active region NOAA~12974 observed with HRI$_{EUV}$ on 30 March 2022 between 00:08\,UT and 00:53\,UT (time measured at Earth) with a cadence of 3\,s and an exposure time of 1.65\,s.
Solar Orbiter was located at 0.33\,AU from the Sun.
The angular separation between Solar Orbiter and the Sun and the Sun-Earth lines was 92$\degree$ in solar longitude, which is ideal for this kind of quadrature view analysis. 
The HRI$_{EUV}$ field of view (FOV) is $2048 \times2048$ pixels, and the plate scale is 118.4\,km at 0.33 AU.

We used level-2 HRI$_{EUV}$ data provided by the EUI team that were
 released as part of the "EUI Data Release 5.0 2022-04". They are publicly  available \citep{euidr5}.

\subsection{Atmospheric Imaging Assembly  observations}
 The AIA continuously observes the full solar disk from the photosphere to the solar corona with seven EUV and three UV channels.
The EUV images are obtained with a cadence of 12\,s, and the UV images are obtained with a cadence of 24\,s.
The pixel size is 0.6\arcsec\ (436\,km) for all AIA channels.

We used pre-processed SDO data provided by the Joint Science Operations Center (JSOC\footnote{\url{http://jsoc.stanford.edu}}). The data were corrected for plate scaling, shifts, and rotation (data are equivalent to level 1.5 data).

We used the images from the AIA~304\,\AA, 171\,\AA, and 193\,\AA\,channels to study the chromosphere, transition region, and the solar corona simultaneously. 
These channels were used to study the  topology and dynamics of the upflow region.

\subsection{Interface Region Imaging Spectrograph observations}
The IRIS is a space-based imaging spectrograph that observes the chromosphere and transition region.
IRIS provides spectroscopic raster data in two far-ultraviolet (FUV)\ channels, at 1332--1358\,\AA\ and 1390--1406\,\AA, and in a near-ultraviolet (NUV) channel at 2786--2835\,\AA.
The spectral sampling is 12.8\,m\AA\ for the FUV and 25.6\,m\AA\ for the NUV.

The IRIS Slit Jaw Imager (SJI) provides the images of the area surrounding the slit in four filters.
Two filters cover the wide bandpass centred at \ion{C}{ii}~1334/1335\,\AA\ and \ion{Si}{iv}~1394/1403\,\AA.
Two other filters focus on the narrower bandpass of 4\,\AA, and they are centred at
\ion{Mg}{ii}~k~2796\,\AA\  and the wing of \ion{Mg}{ii} around 2830\,\AA.

We analysed the active region observed with SJI on 30 March 2022 between 00:23\,UT and 00:53\,UT.
 IRIS observed using a large sit-and-stare mode and provided SJI images in the \ion{Si}{iv}~(1403\,\AA) and \ion{Mg}{ii}~k~(2796\,\AA) filters.
The SJI provides context images of the region around the location under the spectrograph slit.
For each filter, the SJI obtained 488 images with a cadence of 4~s~and an exposure time of 1~s.
The FOV of SJI images is $167\arcsec\times175\arcsec$.
We only analysed the SJI images because we are interested in studying the temporal evolution of our target with a large field of view.
The raster sit-and-stare mode allows us to analyse only single-slit positions.

We downloaded SJI level-2 data from the IRIS database\footnote{\url{https://iris.lmsal.com/search/}}.
These data were reoriented to common axes: north is up (0$^{\circ}$ roll), and they were corrected for dark current and flat field.

\subsection{Hinode EIS observations}
The EIS is a spectrograph that observes the solar atmosphere in two wavelength ranges: 171--211\,\AA\ and 245--291\,\AA.
The spectral resolution in both channels is 22.3\,m\AA.
We analysed the EIS data obtained on 30 March 2022 between 00:32\,UT and 02:59\,UT.
EIS provides spectroscopic data with a slit size of 2\arcsec\ and an FOV of $50.0\arcsec\times152.0\arcsec$.

We downloaded level-0 data from the Hinode/EIS archive\footnote{\url{http://solarb.mssl.ucl.ac.uk/SolarB/SearchArchive.jsp}}.
We reduced the data with the \verb+eis_prep+ standard calibration routine from the Solar Software IDL library.
This routine removed the CCD dark current, cosmic-ray pattern, and hot and dusty pixels from detector exposures and provided the radiometric calibration to physical units (erg cm$^{-2}$ s$^{-1}$ sr$^{-2}$ \AA$^{-1}$).
The data were also corrected for the slit tilt \citep{Warren2014}.

We focused on the strong coronal emission line \ion{Fe}{xii} (195.12\AA).
The \verb+eis_auto_fit+ routine\footnote{Solar Software; eis\_auto\_fit.pro} was used to fit a single Gaussian for the \ion{Fe}{xii} spectral line at each spatial pixel.
There are two spectral lines of the \ion{Fe}{xii} doublet at 195.12\AA\ and 195.18\AA. 
We found the same $\chi
^{2}$ and velocity error for the single- and double-Gaussian fits and hence used the single-Gaussian fit.

We used the fitted parameters to obtain the maps of the peak intensity and Doppler velocity.
The non-thermal velocity map was computed with the solar software \verb+eis_width2velocity+ routine.
This routine uses the equation
$\text{FWHM}^2=(\text{FWHM}_\text{instr})^2+4\ln2\,(\lambda/c)^2(V_t+V_{nt}^2)$, where FWHM is the full width at half maximum,  $\text{FWHM}_{instr}$ is the instrumental full width at half maximum, $\lambda$ is the wavelength of the peak of the emission line, $c$ is the speed of the light, $V_t$  is the thermal velocity, and $V_{nt}$ is the non-thermal velocity.

\begin{figure*}[!htb]
\center{\includegraphics[width=\textwidth]
{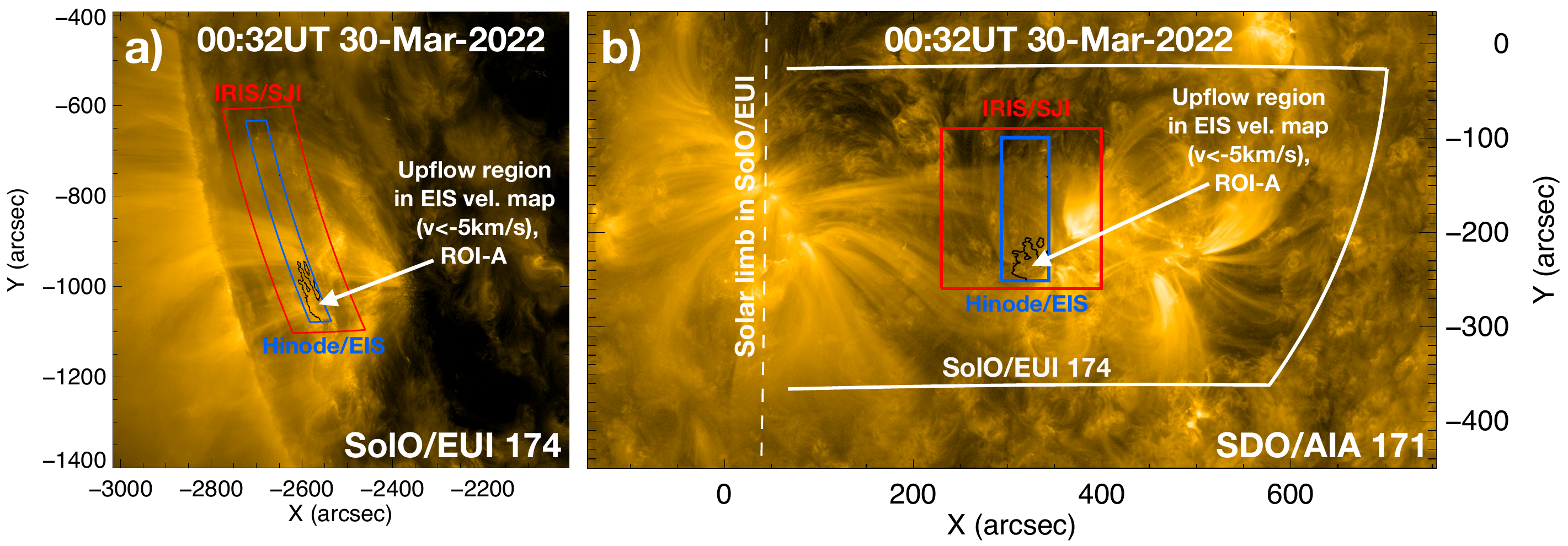}}
\caption{Overview  of  the active region observations from Solar Orbiter/EUI and $HRI_{EUV}$ telescope at 174\,\AA~(a) and AIA~171\,\AA~(b). The IRIS/SJI and Hinode/EIS FOVs are marked with red and blue boxes, respectively. The black contour highlights the ROI-A that spatially corresponds to the Hinode/EIS upflow region defined in Fig.~\ref{fig:hinode_eis}. In panel (b), the white line presents the solar disk limb position from panel (a) computed with the WCS\_LIMB solar software routine and then transformed into the SDO/AIA view. The limb observed in EUI/FSI 174\,\AA\ (panel a) is marked by a dashed line in the ~171\,\AA\ map (panel b).}  \label{fig:overview}
\end{figure*}

\begin{figure}[!htb]
\includegraphics[width=8.8cm]
{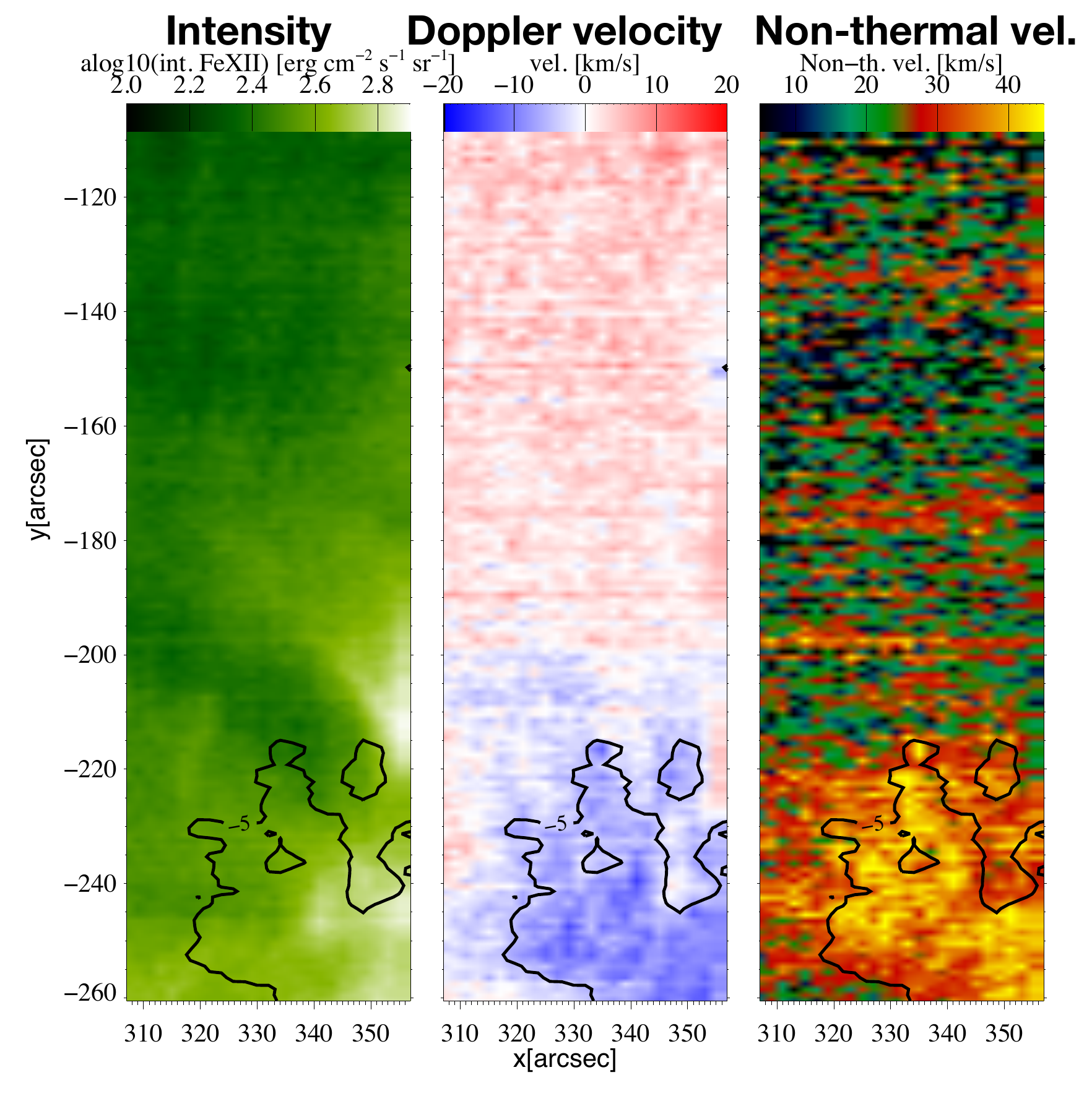}
\caption{Coronal upflow region as observed with Hinode/EIS on 30 March 2022 at 00:32\,UT. The panels show the intensity (left column), Doppler velocity (middle column), and non-thermal velocity (right column) in the \ion{Fe}{xii} emission line (1.0\,MK). Black contours represent a Doppler velocity of $-5$\,km\,s$^{-1}$. An area with Doppler velocities $v<-5$\,km\,s$^{-1}$ is defined as an upflow region.}  \label{fig:hinode_eis}
\end{figure}

\begin{figure*}[!htb]
\includegraphics[width=\textwidth]
{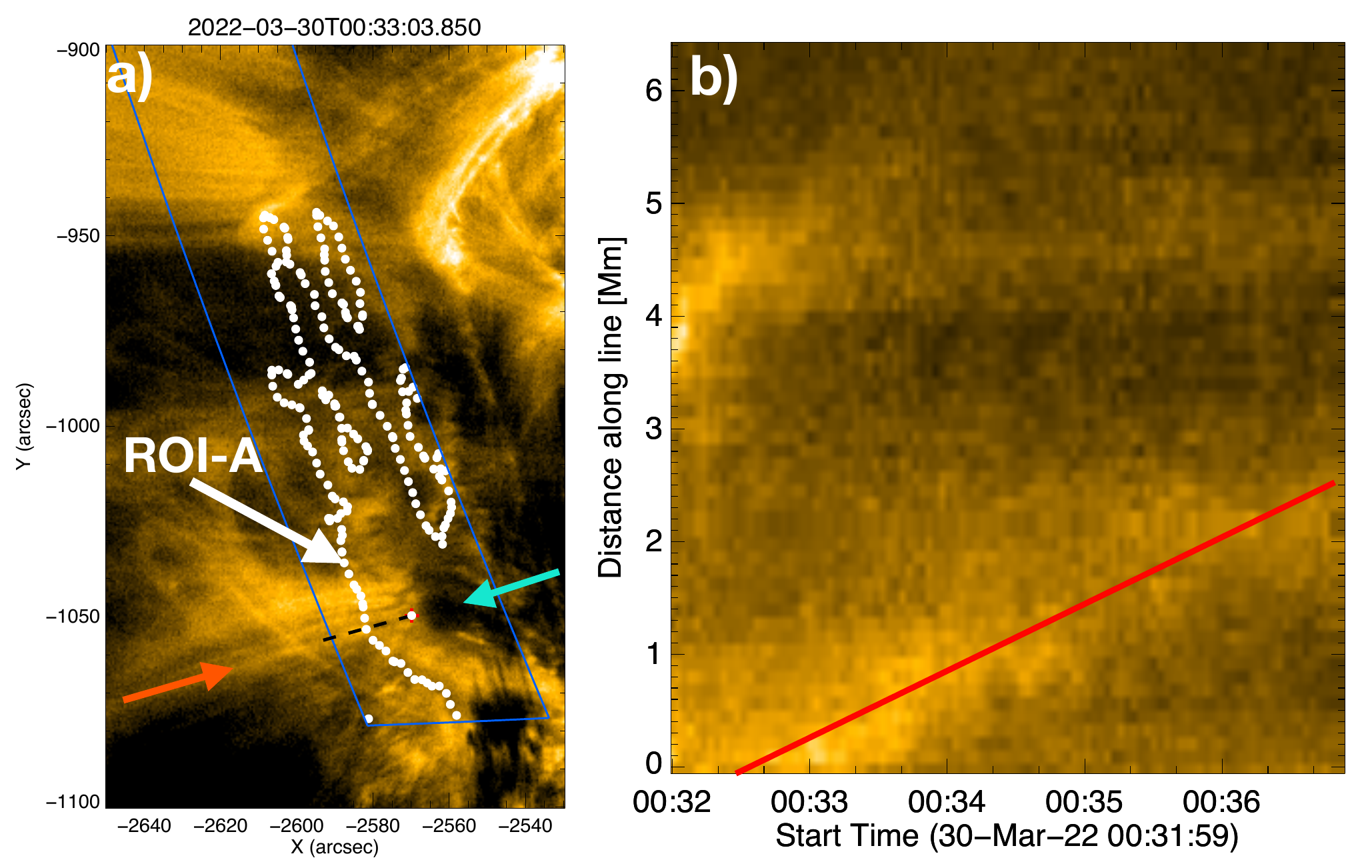}
\caption{
HRI$_{EUV}$ intensity image of the area corresponding to the Hinode/EIS upflow region and its surroundings (Fig.~\ref{fig:hinode_eis}). In panel (a), the white contour highlights the re-projected ROI-A that corresponds to the Hinode/EIS upflow region with Doppler upflow velocities $v<-5$\,km\,s$^{-1}$. The dashed black line highlights the location of the single-pixel-wide line used to measure the velocity along the loops. The extended loop is marked with a red arrow. The blue arrow shows a low-intensity area in the upflow region. Panel (b) shows a change in intensity along the extended loop with time obtained from HRI$_{EUV}$ observations (see the profile line in the figure to the left). The red line is fitted to the locations with the highest intensity and has a velocity of $9.9\pm0.4$\,km\,s$^{-1}$. All panels are scaled to the same intensity range (900-2300 DN/s). The footpoint of the extended loop has a Doppler velocity determined by Hinode EIS of $-8.6\pm2.3$\,km\,s$^{-1}$ (Fig.~\ref{fig:hinode_eis}).}  \label{fig:upflow_vel}
\end{figure*}

\begin{figure*}[!htb]
\includegraphics[width=\textwidth]
{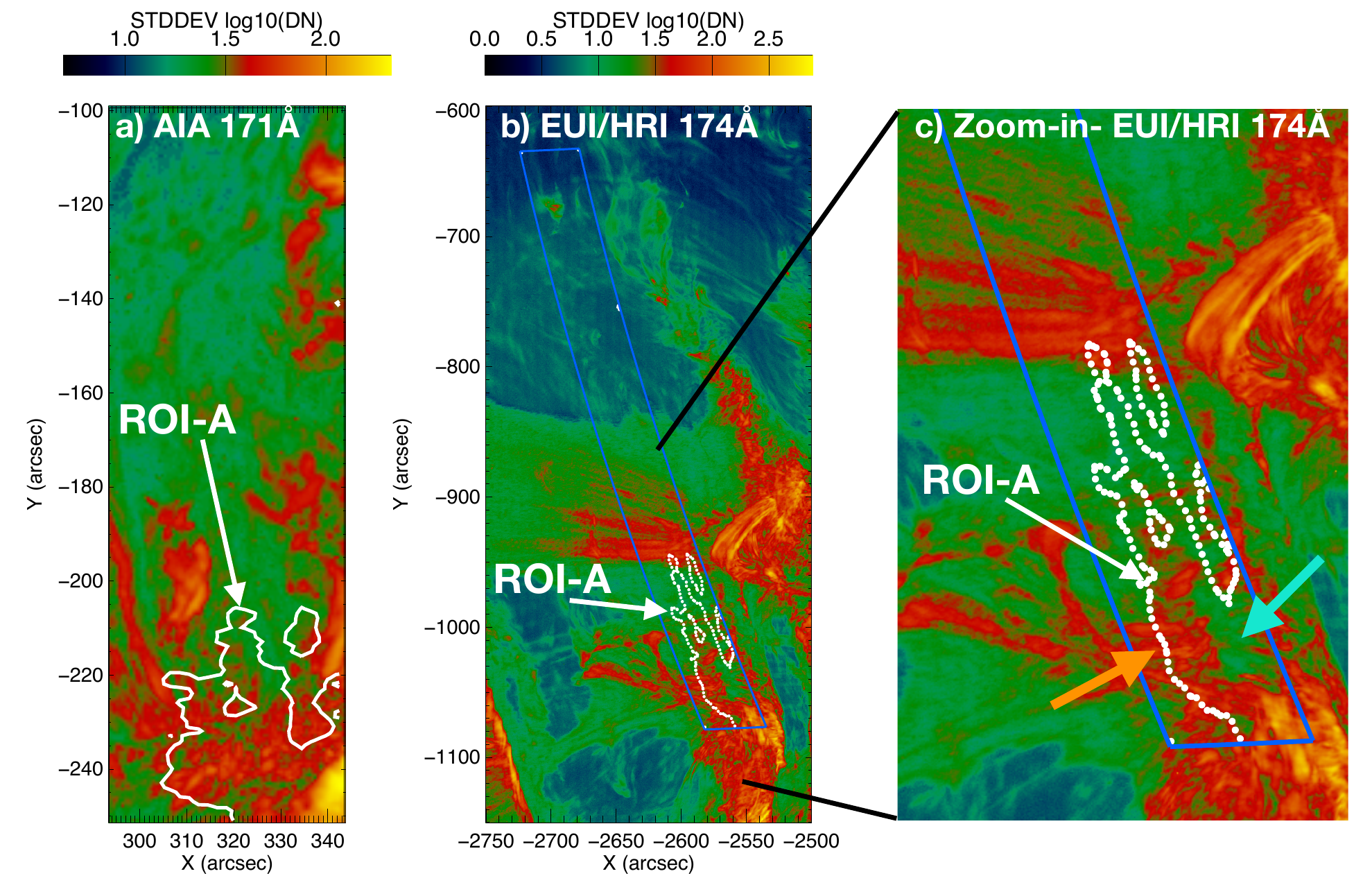}
\caption{Map of the standard deviation of the coronal intensity with time observed with  (a) AIA~171\,\AA\, and (b) HRI$_{EUV}$.   Panel (c) shows a zoom of the upflow region seen by the HRI$_{EUV}$. The maps are computed for the time range between 00:32 to 00:52\,UT on 30 March 2022. The contour highlights the ROI-A that spatially corresponds to the upflow region (velocity $v<-5$\,km\,s$^{-1}$) in Hinode/EIS map (Fig.~\ref{fig:hinode_eis}b). In panel (c), the orange arrow shows the region where there are rooted extended loops (see also Fig. 5) with the standard deviation higher than in the surroundings. The blue arrow shows the region with lower standard deviation intensity values than the extended loops footpoint region. The low standard deviation region corresponds to a low intensity region in the corona maps (see Fig. 5).}\label{fig:stddev0}
\end{figure*}


\begin{figure*}[!htb]
\includegraphics[width=\textwidth]
{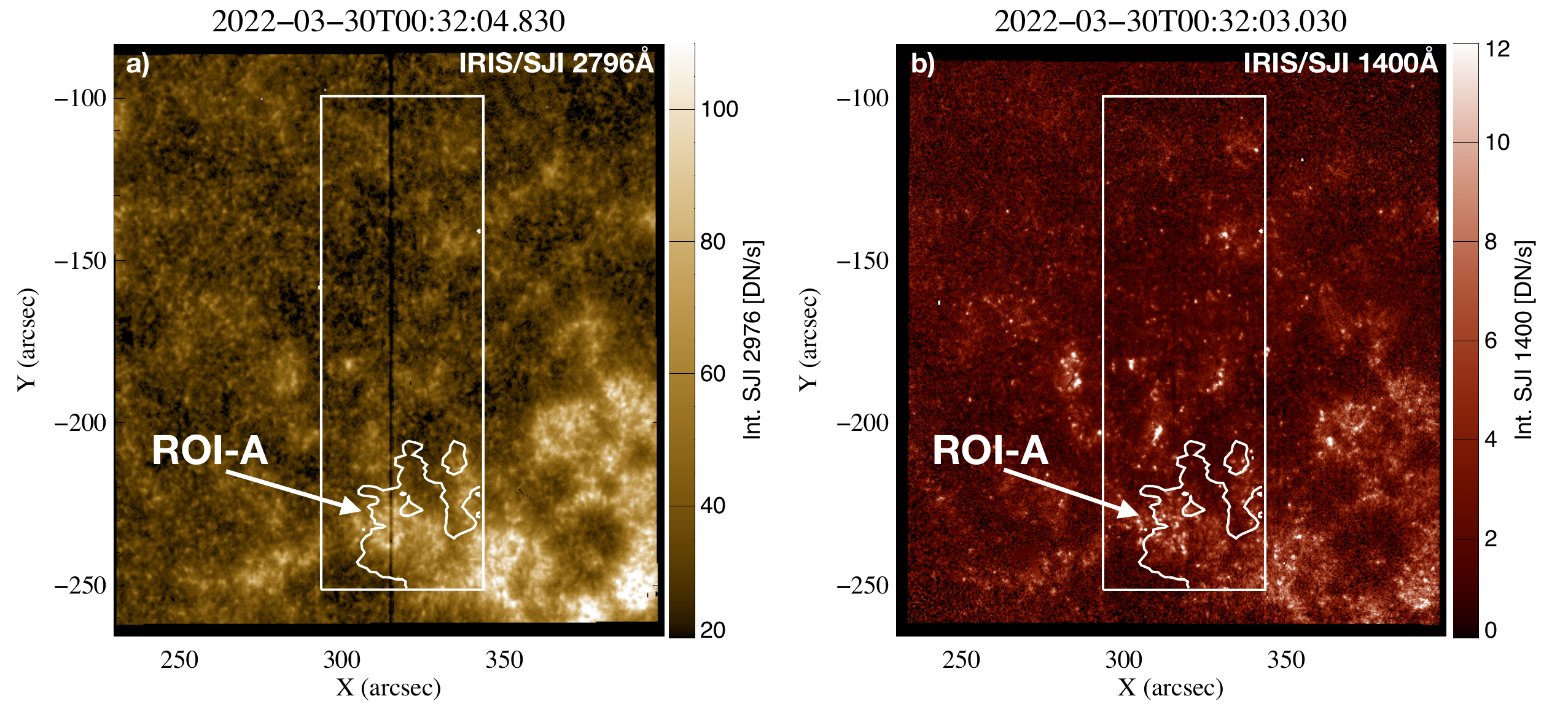}
\caption{
 Highly dynamical small-scale structures are observed in the transition region with IRIS SJI 2796 (panel a) and IRIS SJI 1400 (panel b). The rectangle shows the Hinode/EIS field of view. The contour inside it presents ROI-A that spatially corresponds to the coronal upflow region with a projected velocity $v<-5$\,km\,s$^{-1}$.}  \label{fig:fig_iris}
\end{figure*}


\begin{figure*}[!htb]
\includegraphics[width=\textwidth]
{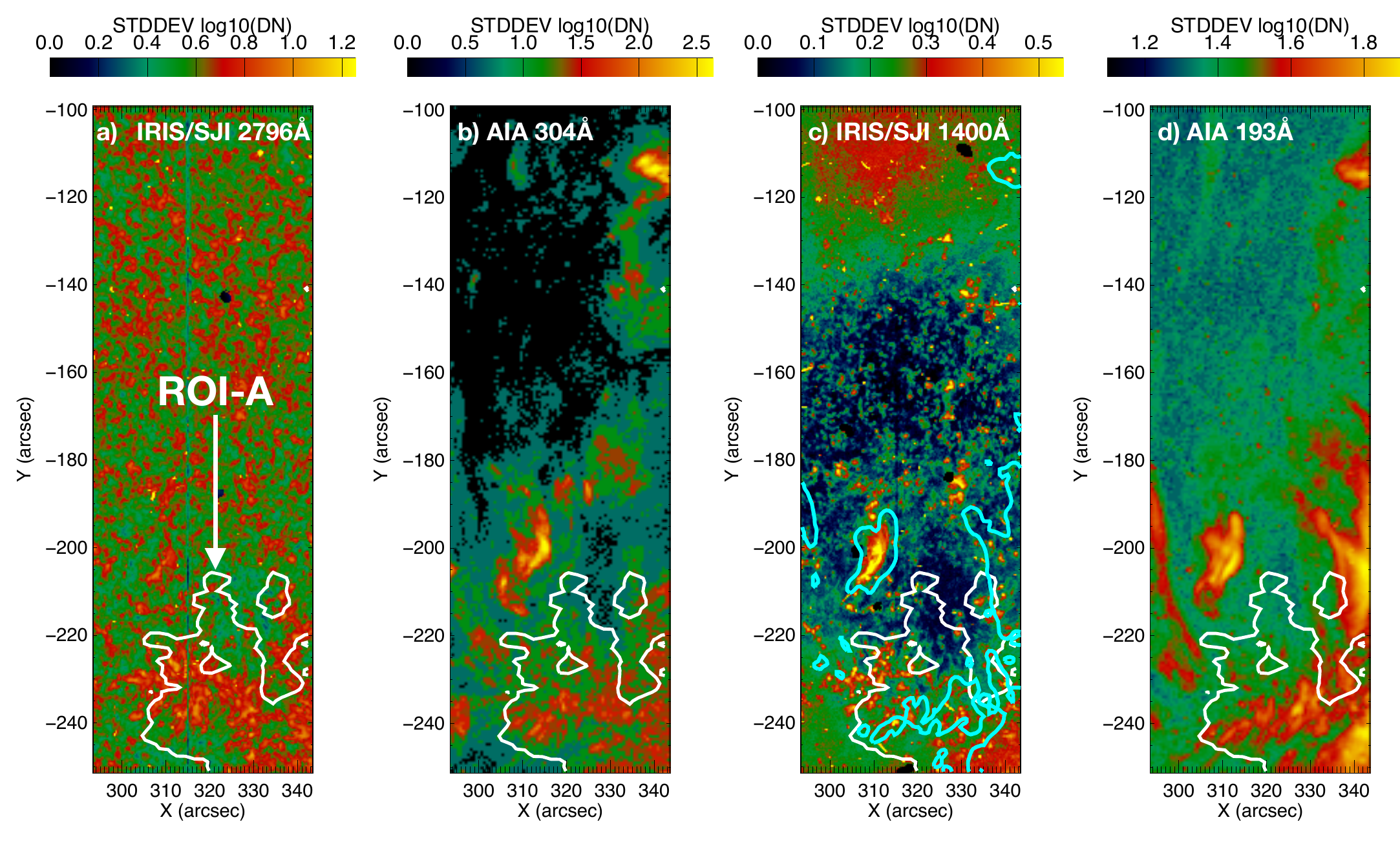}
\caption{
Standard deviation map of the chromosphere (a, b), transition region (c), and the corona (d). The maps are computed for the time range between 00:32 to 00:52\,UT on 30 March 2022. The white contour presents the ROI-A that spatially corresponds to the coronal upflow region with a velocity $v<-5$\,km\,s$^{-1}$. The blue contour in panel (c) shows a strong standard deviation of AIA~193\AA\ intensity of 20 DN/s (panel d).}  \label{fig:stddev}
\end{figure*}

\subsection{Data alignment and reprojection}

We aligned the intensity map of the lines that were formed with similar temperatures using a cross-correlation technique.
We aligned the maps according to the following chain:
SJI~2796\,\AA\ -- SJI~1400\,\AA\ -- AIA~304\,\AA\ -- AIA~171\,\AA\ -- AIA~and 193\,\AA\ -- EIS~\ion{Fe}{xii}.
The accuracy of the alignment between IRIS and AIA data is a single AIA spatial pixel.
The Hinode/EIS and AIA data were aligned with an accuracy of a single Hinode/EIS spatial pixel.

The HRI$_{EUV}$ images were co-aligned together with a sub-pixel precision using a cross-correlation method.
The correction of HRI$_{EUV}$ data was necessary to reduce the effect of the spacecraft jitter.

The re-projection method was used to define the position of the same region of interest (ROI) at the solar disk for an observation with satellites located at a separation angle of 92$\degree$ .
We used the World Coordinate System \citep[WCS;][]{Greisen2002} and solar software WCS routines to re-project the position of the points defined in Hinode/EIS to the corresponding position in HRI$_{EUV}$ maps.

\begin{table}[]
\caption{Spectral lines or passbands of Hinode/EIS, IRIS, EUI, and SDO/AIA used in the analysis. The abbreviations in the "Usage" column define the usage for velocity measurements (vel.) and dynamics of the upflow region substructure studies (dyn.).}
\begin{tabular}{@{}lllll@{}}
\toprule
Line/channel & Wavelength [\AA] & log(T/K)  & Instr. & Usage      \\ \midrule
Fe XII       & 195.12 & 6.2     & EIS    & vel., dyn. \\
AIA~193      & 193    & 6.1     & AIA    & dyn.       \\
EUI 174      & 174    & 5.8        & EUI    & vel., dyn. \\
AIA~171      & 171    & 5.8     & AIA    & dyn.       \\
AIA~304      & 304    &   4.7      & AIA    & dyn.       \\
SJI 1400     & 1400   & 3.7-5.2 & IRIS   & dyn.       \\
SJI 2796     & 2796   & 3.7-4.2        & IRIS   & dyn,       \\ \bottomrule
\end{tabular}
\label{tab:tab1}
\end{table}

\section{Upflow region topology and speed in the quadrature observations}\label{sect:topology}

For the first time,  solar coronal observations obtained simultaneously with SDO/AIA~171\,\AA\ (Fig.~\ref{fig:overview}a) and HRI$_{EUV}$~(Fig.~\ref{fig:overview}b) show an active region with high-spatial and temporal resolutions, and with a large angular separation (92$^{\circ}$).
Based on the EIS Doppler velocity map presented in \ion{Fe}{xii} (Fig.~\ref{fig:hinode_eis}), we defined the upflow region as the area with velocities $v<-5$\,km\,s$^{-1}$.
The upflow region is highlighted with the black contour in the Hinode/EIS maps (Fig.~\ref{fig:hinode_eis}a-c).
We define the region-of-interest A (ROI-A) as the area that spatially corresponds to the upflow velocity $v<-5$\,km\,s$^{-1}$ obtained with Hinode/EIS (Fig.~\ref{fig:hinode_eis}b).
The error associated with the Doppler velocity measurements is related to the fitting, and in the upflow region, it is around $\pm2
$\,km\,s$^{-1}$.
The obtained error is not the absolute error because the Doppler velocities in \ion{Fe}{xii} obtained from EIS lack absolute calibration.
We obtained the error of the velocity measurement using SSW IDL procedure \verb+eis_get_fitdata+.
The intensity map of \ion{Fe}{xii} (Fig.~\ref{fig:hinode_eis}a) shows the absence of the large hot coronal loops in the coronal upflow region from the on-disk view.
The non-thermal velocity map (Fig.~\ref{fig:hinode_eis}c) shows
higher non-thermal velocities in the upflow region than in the surroundings.
This effect is well known and has been observed before \citep[e.g.][]{Chen2011}.

The ROI-A (Fig.~\ref{fig:overview}, black contour) is located at the edge of active region NOAA~12974.
The extended loops are rooted in the ROI-A as is seen from a side view in the HRI$_{EUV}$ map (Fig.~\ref{fig:overview}a).
The face-on view in AIA~171\,\AA\ (Fig.~\ref{fig:overview}b) shows a lower intensity in ROI-A than in its surroundings.

The zoom-in of the HRI$_{EUV}$ observation (Fig~\ref{fig:upflow_vel}a) shows a sub-region with a strong intensity and numerous extended loops rooted in ROI-A.
The HRI$_{EUV}$ imaging data shows ROI-A obtained in quadrature with respect to EIS spectroscopy maps.  
The temporal analysis shows  that brightenings move outward along the loops from the ROI-A centre (see Movie1).
The extended loops (Fig~\ref{fig:upflow_vel}a, red arrow) are bright, and the relative changes in intensity with time are weak. 
To identify the intensity changes, we analysed the movie (movie1) and prepared the stack plot shown in Fig~\ref{fig:upflow_vel}b.

The stack plot (Fig~\ref{fig:upflow_vel}b) shows the intensity changes along the dashed line with time (Fig~\ref{fig:upflow_vel}a).
The dashed line begins at the loop footpoint (red cross in Fig.~\ref{fig:hinode_eis}b).
The beginning of the profile line corresponds to a distance 0\,Mm in the stack plot (Fig~\ref{fig:upflow_vel}c).
The intensity propagation along the reference line is linear with time from 00:32:30\,UT until the end of the stack plot time axis (00:37:00\,UT).
To determine the velocity of this intensity propagation, we defined the maximum intensity point for each time step along the reference line in the distance range 0--3\,Mm.
Then, we fitted a linear function to these points of maximum intensity.
The fitted parameters define the intensity movement along the extended loop due to plasma flow or wave propagation.
The fitting error is interpreted as the error of the velocity measurement.
The increase in intensity propagates with a velocity of $9.6\pm0.4$\,km\,s$^{-1}$ along the extended loops.

For 20\,min of coordinated observations of Hinode/EIS and EUI, we identified five footpoints of extended loops and the plasma flow along these extended loops.
We computed the Doppler velocity from EIS at the footpoints and the velocity along the extended loops from EUI. 
The results of this measurement are summarized in Table~\ref{tab:tab2}.
Measurements 4 and 5 in Table~\ref{tab:tab2} were obtained for the same footpoint and the same reference line position along the extended loop, but at consecutive times.
Thus, they present a different propagation, but originate from the same region.
The Doppler velocity in the upflow region is consistent with previous upflow studies \cite[e.g.][]{Barczynski2021}.
Longer series of data are necessary to describe this propagation and verify their possible periodicity.

The Doppler velocity computed from EIS and the velocity along the extended loops from EUI can be affected by the projection effects.
For the stronger inclination angle of the extended loop from the line normal to the solar surface, the real velocity of the plasma upflow is higher than the measured velocity.
Therefore, the results presented in Table~\ref{tab:tab2} should be interpreted as the lowest velocities.

\begin{table*}[]
\caption{Velocity measurements of the extended loop footpoints and the plasma propagation along the extended loops. The start and end positions (fifth and sixth columns) are measured in the $HRI_{EUV}$ maps in Fig.~\ref{fig:upflow_vel}a}
\begin{tabular}{@{}lllllll@{}}
\toprule
No. & \begin{tabular}[c]{@{}l@{}}Time (Hinode/EIS)\\ 30 March 2022\end{tabular} & \begin{tabular}[c]{@{}l@{}}Position EIS\\ (x, y) {[}arcsec{]}\end{tabular} & \begin{tabular}[c]{@{}l@{}}Vel. EIS\\ {[}km s$^{-1}${]}\end{tabular} & \begin{tabular}[c]{@{}l@{}}Position start\\ (x, y) {[}arcsec{]}\end{tabular} & \begin{tabular}[c]{@{}l@{}}Position end\\ (x, y) {[}arcsec{]}\end{tabular} & \begin{tabular}[c]{@{}l@{}}Vel. EUI\\ {[}km s$^{-1}${]}\end{tabular} \\ \midrule
1   & 00:32:02 - 00:36:55\,UT                                                    & $(319.0, -241.7)$                                                            & $-8.6\pm2.3$                                                          & $(-2569.8,  -1049.7)$                                                          & $(-2596.0, -1057.0)$                                                         & $9.9\pm0.4$                                                           \\
2   & 00:32:02 - 00:36:55\,UT                                                    & $(324.2 , -232.1)$                                                           & $-6.7\pm2.3$                                                          & $(-2576.0, -1021.7)$                                                           & $(-2596.9, -1020.9)$                                                         & $10.1\pm1.5$                                                         \\
3   & 00:36:55 - 00:41:48\,UT                                                    & $(325.2, -238.8)$                                                            & $-9.6\pm2.0$                                                          & $(-2567.4, -1041.0)$                                                           & $(-2655.7, -1072.4)$                                                         & $7.7\pm1.6$                                                           \\
4   & 00:36:55 - 00:41:48\,UT                                                    & $(327.0,  -239.2)$                                                           & $-9.3\pm2.1$                                                          & $(-2565.2, -1041.0)$                                                           & $(-2602.2,  -1048.8)$                                                        & $12.1\pm1.7$                                                          \\
5   & 00:41:48 - 00:46:41\,UT                                                    & $(327.0, -239.2)$                                                            & $-9.0\pm2.0$                                                          & $(-2565.2, -1041.0)$                                                           & $(-2602.2,  -1048.8)$                                                        & $10.1\pm1.7$                                                          \\
6   & 00:46:41 - 00:51:34\,UT                                                    & $(319.4, -238.4)$                                                            & $-12.0\pm2.0$                                                         & $(-2573.6, -1040.0)$                                                           & $(-2594.6, -1045.3)$                                                         & $4.1\pm0.5$                                                           \\ \bottomrule
\end{tabular}
\label{tab:tab2}
\end{table*}

\section{Dynamics of the features in the upflow region}\label{sect:dynamics}

\subsection{Upflow region dynamics with a stereoscopic view}\label{sect:dynamics_stereo}

The quadrature observations show small-scale structures in the EIS upflow regions in the coronal maps of AIA~171\,\AA\ and HRI$_{EUV}$.
We focus on the observations obtained between 00:32\,UT to 00:52\,UT.
These observations show strong intensity changes with time inside ROI-A in the coronal maps obtained by AIA~171\,\AA\ and HRI$_{EUV}$.
We created a spatial map of the standard deviation intensity with time for the AIA~171\,\AA\ (Fig.~\ref{fig:stddev0}a) and HRI$_{EUV}$~(Fig.~\ref{fig:stddev0}b) observations to identify the regions with stronger dynamics.
To do this, we computed the standard deviation of intensity with time separately for each pixel position. We used the IDL procedure STDDEV\footnote{\url{https://www.l3harrisgeospatial.com/docs/stddev.html}} , which calculates the standard deviation as the square root of the variance of intensity with time.
The cadence, exposure time, and instrumental response are different for HRI$_{EUV}$~and AIA~171\,\AA, 
and therefore we can quantitatively compare values within the standard deviation map of either AIA~171\,\AA\ or HRI$_{EUV}$, but not between these two maps.
However, we can qualitatively identify the regions of the stronger and lower standard deviations and compare the position of these regions in the HRI$_{EUV}$~and AIA~171\,\AA\ channels.

The HRI$_{EUV}$~and AIA~171\,\AA\ channels both show an enhanced standard deviation of the intensity fluctuation in ROI-A.
The spatial distribution of the enhanced standard deviation values is different between the instruments. 
In the AIA~171\,\AA\ channel, most of the ROI-A region shows a high standard deviation of the intensity, and only the northern part of ROI-A shows a lower standard deviation of the intensity than the rest.
The HRI$_{EUV}$ map of ROI-A shows a stronger standard deviation in the southern and eastern part of ROI-A (Fig.~\ref{fig:stddev0}c, orange arrow); 
the western part of the ROI-A shows a lower standard deviation (Fig.~\ref{fig:stddev0}c, blue arrow).
The intensity image of HRI$_{EUV}$ (Fig.~\ref{fig:upflow_vel}a) shows many extended loops rooted in the east side of the ROI-A.
The evolution of these extended loops and their footpoints results in a large standard deviation in the eastern and southern part of ROI-A observed in HRI$_{EUV}$.

Even though HRI$_{EUV}$~and AIA~171\,\AA\ show the same region with a similar peak temperature response, the difference in the spatial distribution of the standard deviation of intensity in the HRI$_{EUV}$ and AIA~171\,\AA\ channels is clearly visible.
We suggest that this difference between HRI$_{EUV}$ and AIA~171\,\AA\ is likely to be related to a projection effect.
The upflow region observations from the side view and face-on view show corresponding structures, but observed from different observation angles.
Thus, the line of sight is different for the side view and for the face-on view, and this implies the difference in intensity integrated along the line of sight.
The temporal evolution of the intensity observed from different orientation angles is different as well, which implies a difference in the standard deviation maps of the intensity for the HRI$_{EUV}$ and AIA~171\,\AA\ observations.


\subsection{Upflow region dynamics in the chromosphere, transition region, and solar corona}\label{sect:dynamics_layers}

The chromosphere (Fig.~\ref{fig:fig_iris}a), transition region (Fig.~\ref{fig:fig_iris}b), and (Fig.~\ref{fig:upflow_vel}) coronal intensity maps  of ROI-A show numerous fast-evolving small-scale structures.
The attached movies of the chromosphere (movie2), transition region (movie3), and the solar corona (movie4) show the evolution of these structures over a period of 20 min.
We identified the sub-regions of ROI-A with high and low concentrations of these small-scale structures.
We found these sub-regions in the chromosphere, transition region, and coronal maps (movies1-4).
We suggest that it is possible to localise places with mechanism I (reconnection of the hot coronal loops with open magnetic field lines in the solar corona) and II (reconnection of the
small chromospheric loops with open magnetic field lines in the chromosphere/transition region) based on the standard deviation maps.
The reconnection of the open magnetic field lines with the closed hot loops in the lower corona or upper transition region
should create a strong increase in intensity and higher dynamics of
the coronal structures, and it is expected to generate strong plasma flows therein, and hence a stronger standard deviation in these parts of the solar atmosphere.
In a similar way, the reconnection in the chromosphere should increase the standard deviation of the intensity in the map of the chromospheric temperatures. 
The strong standard deviation of the intensity in the lower corona/upper transition region suggests mechanism I, that is, reconnection of the hot coronal loops with open magnetic field lines in the solar corona. The strong standard deviation of the intensity in the chromosphere suggests mechanism II, that is, reconnection of the small chromospheric loops with open magnetic field lines in the chromosphere/transition region.

We created spatial maps of the standard deviation of the intensities with time of the chromosphere (Fig~\ref{fig:stddev}a, b), transition region (Fig~\ref{fig:stddev}c), and the corona (Fig~\ref{fig:stddev}d).
The evolution in the maps of the standard deviation of the intensity with time helps identify the places with the strongest intensity changes.
In all of the analysed channels, ROI-A shows a stronger standard deviation of the intensity than in the quiet Sun.

The standard deviation map of the intensity in the solar corona shows stronger fluctuations in the centre of ROI-A than in the surroundings.
In the transition region and chromosphere, we found places with a high standard deviation that corresponded to the place in the corona with a high standard deviation of the intensity.
The blue contour highlights the strong coronal (AIA\,193; Fig~\ref{fig:stddev}d) standard deviation of the intensity in the transition region map  (Fig~\ref{fig:stddev}c).
We suggest that the places with the strong standard deviation of the intensity in the transition region map that are located beyond the blue contour, but still in the region that spatially corresponds to coronal upflow region (white contour), can show a different upflow mechanism than the region inside the blue contour.
This is because the process that causes the intensity evolution in the region highlighted by the blue contour occurs in the solar corona; for instance,\ reconnection of hot coronal loops with open magnetic field features.
The high standard deviation that only occurs in the transition region map suggests that the process that causes the change in intensity is located in this layer of the solar atmosphere. It might be\ the reconnection between the small chromospheric/transition region loops with the open magnetic field lines, for example.
Further analysis of similar observations is required to better understand the upflow mechanisms.

\section{Discussion}\label{sect:discussion}
We analysed an active region that was for the first time observed in a quadrature view with data that have a high spatial and temporal resolution.
To do this, we used imaging data obtained from AIA~171\,\AA\ on board the Solar Dynamics Observatory and HRI$_{EUV}$ on board Solar Orbiter.
Moreover, we used simultaneous spectroscopic observations obtained by IRIS and Hinode/EIS.
Hinode/EIS observed plasma at coronal temperatures and allowed us to investigate the plasma flow.
We defined the upflow region as a compact area with Doppler velocities $v<-5$\,km\,s$^{-1}$ in the \ion{Fe}{xii} line observed with Hinode/EIS. 
In the coronal maps of the HRI$_{EUV}$ intensity, we found extended loops rooted in the area that corresponds to the upflow region in the Hinode/EIS map.
The plasma flow measured along the extended loop with HRI$_{EUV}$ is similar to the Doppler velocities at the footpoint of these loops, as shown in the Hinode/EIS Doppler velocity maps.
Then, we investigated the dynamics of the coronal upflow region and spatially corresponding underlying layers of the solar atmosphere.
The high-resolution observations obtained with IRIS and AIA show numerous small-scale structures in the chromosphere and transition region in the locations that correspond to the coronal upflow region observed with Hinode/EIS.
The standard deviation maps allowed us to distinguish the regions where mechanism I (reconnection between the hot coronal loop and open magnetic field) and mechanism II  (reconnection between the chromospheric small-loop and open magnetic field) play a dominant role.

Here, we summarise the results of our study. From a limb-view, the high-resolution quadrature observation of the active region shows extended loops that are rooted in the area that corresponds to the upflow region in the coronal Doppler velocity map (Section~\ref{sect:upflow_topology}). The coronal footpoints of the extended loops show upflow Doppler velocities (\ion{Fe}{xii} line, Hinode/EIS) that are similar to plasma velocities that are measured along the expanded coronal loops from the side view (Solar Orbiter, HRI$_{EUV}$) (Section~\ref{sect:plasma_flow-meas}). The intensity maps in the area that spatially corresponds to the coronal upflow region shows strong dynamics in the small-scale structures in the chromosphere, TR, and corona (Section~\ref{sect:plasma_flow-meas}). Solar observations with two instruments in quadrature with respect to the Sun (HRI$_{EUV}$ vs.\ AIA~171) show different variations. This might be caused by the different lines of sight in the corona (Section~\ref{sec:dynamics_ssf}).

\subsection{Upflow region topology in the quadrature observation of the solar disk}\label{sect:upflow_topology}
Our quadrature observations of an active region show extended loops in the corona that are rooted in the place that corresponds to the coronal upflow region in the Hinode/EIS map (Section~\ref{sect:topology}).
The imaging observations suggest that the extended loops are related to the upflow region, but the projection effect can strongly affect a proper determination of the position of long and expanded features such as loops.
Therefore, the high-resolution simultaneous imaging observation of the active region is important to confirm that the extended loops are rooted in the region that spatially corresponds to the upflow region observed in the solar corona with Hinode/EIS.
Moreover, a negative magnetic field polarity is always connected to a positive magnetic field polarity because magnetic monopoles do not exist. The so-called open magnetic field lines remain attached to two opposite polarities at the solar surface and expand into the heliosphere. They are therefore described as open magnetic field lines. The extended loops are similar to so-called open magnetic field flux tubes. The extended loops can extend significantly beyond the solar corona, and hence they can contribute to the slow solar wind as well.
This implies that the plasma can propagate from the upflow region along the extended loops.
This result is consistent with previous studies of the magnetic field topology of the upflow region \citep[e.g.][]{Baker2017}.

\subsection{Plasma upflow measurements in the quadrature observation}\label{sect:plasma_flow-meas}
The Doppler velocity map of the upflow region shows the velocity in the range between $-20$\,km\,s$^{-1}$ and $-5$\,km\,s$^{-1}$.
We selected five footpoints of the expanded coronal loops observed with HRI$_{EUI}$ rooted in the area corresponding to the upflow region observed with Hinode/EIS.
We measured the plasma Doppler velocity at these footpoints and the velocity along extended loops that are rooted at the footpoints at which we measured the Doppler velocity.
We obtained similar velocities of about 10\,km\,s$^{-1}$ for the plasma upflow in footpoints and for a plasma propagation along the extended loops.
The similar results obtained from two different methods of the upflow plasma measurements suggest that the same plasma flow was measured from the on-disk and close-to-limb view.

Our results are the first quadrature observation measurements of the plasma velocity in an active region upflow region.
Further analysis based on a larger number of upflow regions is necessary to provide a general view of the upflow region plasma flow measured from quadrature observations.
This method should be applied to the data from the Spectral Imaging of the Coronal Environment \citep[SPICE;][]{Spice2020} imaging spectrometer on board Solar Orbiter to provide the first-ever quadrature-spectral observation of an upflow region \citep{Podladchikova2021}.
SPICE observed this upflow region on 30 March 2022 at 00:00 UT.
This SPICE data analysis will largely depend on how well the SPICE spectra can be corrected for the point spread function to retrieve reliable Doppler shifts \cite[][]{Plowman2022}. 

\subsection{Dynamics of the small-scale features in the area that corresponds to the coronal upflow region}\label{sec:dynamics_ssf}

The chromosphere, transition region, and solar corona maps show numerous small-scale features in the area that spatially corresponds to the coronal upflow region.
The standard deviation maps allowed us to identify a sub-region with a stronger intensity fluctuation than its surroundings in the area that spatially corresponds to the coronal upflow region.

\citet{Barczynski2021} discussed three mechanisms that can generate plasma upflows.
We suggest that mechanism I, the reconnection of the open magnetic field lines with the closed hot loops in the lower corona or upper transition region, probably creates a strong increase in intensity and in the higher dynamics of the coronal structures, and generates strong plasma flows in the solar corona.
Therefore, we expect that mechanism I creates a higher standard deviation of the intensity closer to the reconnection location in the solar corona than far from the reconnection location.
Mechanism II is reconnection between open magnetic field lines with small-scale loops in the chromosphere and the lower transition region \citep{Barczynski2021}. 

The region with the stronger standard deviation of the intensity identified in the transition region but with a low standard deviation of the intensity in the solar corona suggests mechanism II of the plasma upflow.
IRIS provides the Doppler velocity maps of the chromosphere and transition region with two modes: sit-and-stare, and raster scan.
To compare Hinode/EIS and IRIS Doppler velocity maps of an upflow region, we need use simultaneous raster-scan-mode observation.
However, IRIS worked in sit-and-stare mode during the quadrature observation. This prevents a comparison of IRIS and Hinode/EIS velocity maps.

\section{Conclusions}\label{sect:conclusion}
We presented the first high-resolution quadrature observations of the solar corona in the region that corresponds to the plasma upflow at the edge of an active region measured in the Doppler velocity map of \ion{Fe}{xii} at 195\AA. Our observations show that the extended loops are rooted in the coronal upflow region. 
These loops were observed before in coronal limb observations, but not in on disk-observations.
Our observations show that blobs of plasma flow from the upflow region along the extended features.

The coronal images of the solar disk near the limb show that the plasma flows along these extended loops have a similar velocity as the plasma upflow velocity measured with the Doppler method at the footpoints of the these loops from the on-disk view. 
This implies that the mechanism that generates plasma upflows at the border of the upflow region might cause the plasma propagation along extended loops.
Moreover, extended loops can extend significantly beyond the solar corona, and hence they can contribute to the slow solar wind as well.
Further analysis based on simultaneous high-resolution quadrature observations of the active region are necessary to better understand the mechanism that causes the plasma upflow at the active region borders.

 The dynamics of small-scale structures in the upflow region can be used to identify the mechanisms driving the plasma upflow.
 Based on standard deviation maps of the intensity with time from different layers of the solar atmosphere, it is possible to distinguish between reconnection of the hot coronal loops with open magnetic field lines in the solar corona (mechanism I) and reconnection of the small chromospheric loops with open magnetic field lines in the chromosphere/transition region (mechanism II).
 Further analysis based on simultaneous high-resolution quadrature observations of the active region are necessary to better understand the mechanism that causes the plasma upflow at the active region borders.

\begin{acknowledgements}
Solar Orbiter is a space mission of international collaboration between ESA and NASA, operated by ESA. The EUI instrument was built by CSL, IAS, MPS, MSSL/UCL, PMOD/WRC, ROB, LCF/IO with funding from the Belgian Federal Science Policy Office (BELSPO/PRODEX PEA 4000134088, 4000112292, 4000136424, and 4000134474); the Centre National d’Etudes Spatiales (CNES); the UK Space Agency (UKSA); the Bundesministerium für Wirtschaft und Energie (BMWi) through the Deutsches Zentrum für Luft- und Raumfahrt (DLR); and the Swiss Space Office (SSO).
SDO data are courtesy of NASA/SDO and the AIA, EVE, and HMI science teams. IRIS is a NASA small explorer mission developed and operated by LMSAL with mission operations executed at NASA Ames Research Center and major contributions to downlink communications funded by ESA and the Norwegian Space Centre. Hinode is a Japanese mission developed and launched by ISAS/JAXA, with NAOJ as domestic partner and NASA and STFC (UK) as international partners. It is operated by these agencies in cooperation with ESA and NSC (Norway).
The funding by CNES through the MEDOC data and operations center.
D.M.L. is grateful to the Science Technology and Facilities Council for the award of an Ernest Rutherford Fellowship (ST/R003246/1). The ROB team thanks the Belgian Federal Science Policy Office (BELSPO) for the provision of financial support in the framework
of the PRODEX Programme of the European Space Agency (ESA) under contract numbers 4000134474, 4000134088, and 4000136424.
\end{acknowledgements}

%
%

\bibliographystyle{aa} 
\bibliography{references.bib} 
\end{document}